# Dynamical Casimir –Polder atom –surface interaction

G.V.Dedkov[1] and A.A.Kyasov

Nanoscale Physics Group, Kabardino –Balkarian State University, Nalchik, 360004

Russian Federation

We have calculated dynamical Casimir –Polder interaction force between a moving ground state atom and a flat polarizable surface. The velocity of an atom can be close to the velocity of light. The material properties are taken into account using a single oscillator model of the atomic dynamic polarizability and the Drude dielectric function of a metal substrate. The limit cases of nonrelativistic velocities and an ideal metal substrate are also considered. We have found specific dependence of the calculated forces on the velocity (energy), distance and material properties.

## 1.Introduction

Starting from pioneering studies [1-4] on conservative van der Waals and Casimir –Polder –Lifshitz forces, the corresponding problem, despite its fundamental nature, has been mainly attacked as a static problem, assuming a zero relative velocity of interacting bodies. First nonrelativistic calculation of the dynamical van der Waals force between a moving atom and the solid surface has been done by Ferrell and Ritchie [5]. An attempt of relativistic calculation has been done by Marvin and Toigo [6], using conventional free –energy approach. However, the used recipe $\omega \to \omega \pm \mathbf{kV}$ ($\omega$ is the frequency and $\mathbf{kV}$ is the Doppler shift) in the expression for free energy of the system “moving atom –surface”, turns out to be incorrect in general case of arbitrary velocity **V**, as we aim to discuss elsewhere [7].

Since 1980’s, due to the lack of consistent relativistic theory of the fluctuation –electromagnetic interaction, considering relative motion, retardation effects, material properties and thermal state of interacting bodies, a straightforward calculation of the dynamical Casimir –Polder force was absent. This work is a natural extension of our recent work [8] in the case of relativistic velocities of atoms with allowance for retardation, using the general formalism developed in [9]. A very comprehensive survey of the past and recent advances made by other authors in the enormously extensive field of van der Waals and Casimir interactions, one can find in the recent resource letter of Milton [10] (see also [11,12]).

[1] Corresponding author e-mail: gv_dedkov@mail.ru

The paper is organized as follows. In the beginning of Section 2, we recall our basic formula from [9] for the dynamical fluctuation electromagnetic force of attraction between a particle of arbitrary velocity and a thick plate (cavity wall), using a real frequency representation. In what follows, we adopt this formula to the case of zero temperature of a system which consists of the ground state atom and a plate. It is worth noting that it is difficult to obtain the foregoing results if the starting formula (3) does not include the particle and substrate temperatures ($T_1$ ,$T_2$). The next part of this section is devoted to analytical analysis of several limit cases corresponding to small velocities of an atom ($\beta = V/c \to 0$, $c$ is the speed of light in vacuum), large velocities ($\beta \to 1$), and the case of perfectly reflecting surface. The Drude case for the surface material proves to be a much more complicated task. The involved formulas in Section 2.3 are reduced to double integrals, allowing to make numerical calculations in Section 3. Conclusions are given in Section 4.

## 2. Theory

Our approach is based on the dipole approximation of fluctuation electromagnetic theory. We consider a case where a small neutral particle (an atom) moves adiabatically at a constant velocity $\mathbf{V}$ in vacuum near the smooth surface of a medium, or equivalently – a thick plate (Fig.1). To calculate the force of attraction the particle to the surface, direct quantum-statistical averaging of the electromagnetic Lorentz force is performed

$$\mathbf{F} = \left\langle \int \left( \rho \mathbf{E} + \frac{1}{c} \mathbf{j} \times \mathbf{B} \right) d^3 r \right\rangle \tag{1}$$

where $\rho$ and $\mathbf{j}$ are the local charge and current densities in the bulk of the particle, $\mathbf{E}, \mathbf{B}$ are the components of fluctuation electromagnetic field, satisfying the Maxwell equations and the appropriate boundary conditions. Assuming the dipole approximation, it is straightforward matter to rewrite Eq.(1) in a more convenient form [9]

$$F_z = \left\langle \nabla_{\mathbf{z}} (\mathbf{d}^{sp} \mathbf{E}^{ind} + \mathbf{d}^{ind} \mathbf{E}^{sp} + \mathbf{m}^{sp} \mathbf{B}^{ind} + \mathbf{m}^{ind} \mathbf{B}^{sp}) \right\rangle \tag{2}$$

where "*sp*" and "*ind*" indicate spontaneous and induced components of electromagnetic field and dipole moments (electric and magnetic) of the particle. Despite being rather tedious, the foregoing calculation in Eq.(2) is carried out explicitly and leads to the formula [9] (for simplicity, we retain hereafter only the terms related with electric polarizability of the particle)

$$F_z = -\frac{\hbar\gamma}{2\pi^2}\int_0^{\infty} d\omega \int_{-\infty}^{+\infty} dk_x \int_{-\infty}^{+\infty} dk_y \left\{ \begin{array}{l} \alpha_e''(\gamma\omega^+)\,\mathrm{Re}\left[\exp(-2q_0 z)R_e(\omega,\mathbf{k})\right]\coth\left(\dfrac{\gamma\hbar\omega^+}{2k_B T_1}\right) + \\ +\,\alpha_e'(\gamma\omega^+)\,\mathrm{Im}\left[\exp(-2q_0 z)R_e(\omega,\mathbf{k})\right]\coth\left(\dfrac{\hbar\omega}{2k_B T_2}\right) \end{array} \right\} \quad (3)$$

$$\begin{gathered} R_e(\omega,\mathbf{k}) = \Delta_e(\omega)\left[2(k^2 - k_x{}^2\beta^2)(1-\omega^2/k^2c^2) + (\omega^+)^2/c^2\right] + \\ +\,\Delta_m(\omega)\left[2k_y{}^2\beta^2(1-\omega^2/k^2c^2) + (\omega^+)^2/c^2\right] \end{gathered} \quad (4)$$

$$\begin{gathered} R_m(\omega,\mathbf{k}) = \Delta_m(\omega)\left[2(k^2 - k_x{}^2\beta^2)(1-\omega^2/k^2c^2) + (\omega^+)^2/c^2\right] + \\ +\,\Delta_e(\omega)\left[2k_y{}^2\beta^2(1-\omega^2/k^2c^2) + (\omega^+)^2/c^2\right] \end{gathered} \quad (5)$$

$$\begin{gathered} \Delta_e(\omega) = \frac{q_0\varepsilon(\omega) - q}{q_0\varepsilon(\omega) + q},\ \Delta_m(\omega) = \frac{q_0\mu(\omega) - q}{q_0\mu(\omega) + q},\ q = \left(k^2 - (\omega^2/c^2)\varepsilon(\omega)\mu(\omega)\right)^{1/2}, \\ q_0 = (k^2 - \omega^2/c^2)^{1/2},\ k^2 = k_x{}^2 + k_y{}^2,\ \omega^+ = \omega + k_x V \end{gathered} \quad (6)$$

where $\alpha_e(\omega)$ is the particle electric polarizability, one –primed and double –primed quantities denote real and imaginary parts, $\varepsilon(\omega)$ and $\mu(\omega)$ denote the bulk dielectric and magnetic permittivities of the plate. In what follows $\mu(\omega) = 1$. The global system is assumed to be out of thermal equilibrium but in a stationary regime, the plate and surrounding vacuum background are assumed to be at equilibrium with temperature $T_2$. Similarly to [8], making use the limit transitions

$$\lim_{T_1\to 0}\coth\frac{\hbar(\omega + k_x V)}{2kT_1} = sign(\omega + k_x V)\,,\ \lim_{T_1\to 0}\coth\frac{\hbar(\omega + k_x V)}{2kT_1} = sign(\omega + k_x V)\,,$$

one can rewrite Eq.(3) as a sum of two terms, of which the first one is represented as integral over imaginary frequencies, and the second –as integral over real frequencies:

$$F_z = F_z{}^{(0)} + F_z{}^{(1)} \quad (7)$$

$$\begin{gathered} F_z{}^{(0)} = -\frac{\hbar\gamma}{2\pi^2}\int_{-\infty}^{+\infty} dk_x \int_{-\infty}^{+\infty} dk_y \cdot \mathrm{Im}\left\{\int_0^{\infty} d\xi \exp\left(-2\sqrt{k^2 + \xi^2/c^2}\,z_0\right)\alpha\left(\gamma(\mathrm{i}\xi + k_x V)\right)\cdot \right. \\ \left. \cdot\left[\mathrm{i}\cdot R_e{}^{(1)}(\mathrm{i}\xi, k) - 2\beta k_x \frac{\xi}{c}\left(\Delta_e(\mathrm{i}\xi)\right) + \Delta_m(\mathrm{i}\xi)\right]\right\} \end{gathered} \quad (8)$$

$$F_z^{(1)}=\frac{2\hbar\gamma}{\pi^2}\int_0^{\infty}dk_x\int_0^{\infty}dk_y\int_0^{k_xV}d\omega\,\alpha''(\gamma(\omega-k_xV))\cdot\mathrm{Re}\Big\{\exp\Big(-2\sqrt{k^2-\omega^2/c^2}\,z_0\Big)\cdot$$
$$\cdot\left[R_e^{(1)}(\omega,k)-2\beta k_x\frac{\omega}{c}\big(\Delta_e(\omega)+\Delta_m(\omega)\big)\right]\Big\} \tag{9}$$

$$R_e^{(1)}(\mathrm{i}\xi,k)=\Delta_e(\mathrm{i}\xi)(2k^2+\xi^2/c^2)+\Delta_m(\mathrm{i}\xi)\left[2\beta^2(k^2+\xi^2/c^2)-\xi^2/c^2\right]-$$
$$-\beta^2\left[\Delta_e(\mathrm{i}\xi)+\Delta_m(\mathrm{i}\xi)\right](k^2+2\xi^2/c^2)\cos^2\theta \tag{10}$$

$$R_e^{(1)}(\omega,k)=(\omega^2/c^2+k^2\beta^2\cos^2\theta)\left[\Delta_e(\omega)+\Delta_m(\omega)\right]+2(k^2-\omega^2/c^2)\cdot$$
$$\cdot\left[(1-\beta^2\cos^2\theta)\Delta_e(\omega)+\Delta_m(\omega)\beta^2\sin^2\theta\right] \tag{11}$$

where $\cos\theta=k_x/k$, $\beta=V/c$.

To make a step further, we employ a single oscillator model of the atomic polarizability

$$\alpha(\gamma\,\omega)=\frac{\alpha(0)\tilde{\omega}_0^{\,2}}{\tilde{\omega}_0^{\,2}-\omega^2-\mathrm{i}\cdot 0\cdot\omega}\;,\;\tilde{\omega}_0=\omega_0/\gamma \tag{12}$$

$$\alpha''(\gamma\,\omega)=\frac{\pi\,\alpha(0)\tilde{\omega}_0}{2}\left[\delta(\omega-\tilde{\omega}_0)-\delta(\omega+\tilde{\omega}_0)\right], \tag{13}$$

where $\omega_0$ is the atomic transition frequency. Substitution into Eqs.(8),(9) via Eqs.(12),(13) yields

$$F_z^{(0)}=-\frac{1}{2\pi^2}\hbar\,\alpha(0)\tilde{\omega}_0^{\,2}\gamma\int_{-\infty}^{+\infty}dk_x\int_{-\infty}^{+\infty}dk_y\int_0^{\infty}d\xi\,\frac{\exp(-2\sqrt{k^2+\xi^2/c^2}\,z_0)}{\left[(\tilde{\omega}_0^{\,+})^2+\xi^2\right]\left[(\tilde{\omega}_0^{\,-})^2+\xi^2\right]}\cdot$$
$$\cdot\left[R_e^{(1)}(\mathrm{i}\xi,k)\left(\tilde{\omega}_0^{\,+}\tilde{\omega}_0^{\,-}+\xi^2\right)-4\beta^2k_x^{\,2}\xi^2\left(\Delta_e(\mathrm{i}\xi)+\Delta_m(\mathrm{i}\xi)\right)\right] \tag{14}$$

$$F_z^{(1)}=-\frac{1}{\pi}\hbar\alpha(0)\omega_0\int_{\tilde{\omega}_0/V}^{\infty}dk_x\int_0^{\infty}dk_y\,\mathrm{Re}\Big\{\exp(-2\sqrt{k^2-(\tilde{\omega}_0^{\,-})^2/c^2}\,z_0)\cdot$$
$$\cdot\left[R_e^{(1)}(-\tilde{\omega}_0^{\,-},k)+2k_x\beta(\tilde{\omega}_0^{\,-}/c)\left(\Delta_e(-\tilde{\omega}_0^{\,-})+\Delta_m(-\tilde{\omega}_0^{\,-})\right)\right]\Big\} \tag{15}$$

where $\tilde{\omega}_0^{\,\pm}=\tilde{\omega}_0\pm k_xV=\tilde{\omega}_0\pm kV\cos\theta$.

Using Eqs.(14) and (15), it is worthwhile to examine several important cases.

*2.1 Nonrelativistic velocities and no retardation*

In the limit $c \to \infty$, Eqs.(14),(15) are simplified to

$$F_z = -\frac{\hbar}{\pi^2}\int_{-\infty}^{+\infty} dk_x \int_{-\infty}^{+\infty} dk_y k^2 \exp(-2k z_0) \cdot \mathrm{Im}\left[ \mathrm{i} \int_0^{\infty} d\xi \, \Delta(\mathrm{i}\xi)\alpha(\mathrm{i}\xi + k_x V) \right] + \\ + \frac{4\hbar}{\pi^2}\int_0^{\infty} dk_x \int_0^{\infty} dk_y k^2 \exp(-2k z_0) \int_0^{k_x V} d\omega \Delta'(\omega)\alpha''(\omega - k_x V), \tag{16}$$

where $\Delta(\omega) = (\varepsilon(\omega) - 1)/(\varepsilon(\omega) + 1)$. Bearing in mind relation $F_z = -\partial U / \partial z$, where $U(z)$ is the van der Waals energy, Eq.(16) is identical to our nonrelativistic result [8].

Making use the dimensionless variables $x = 2z_0 k_x$, $y = 2z_0 k_y$, $p = \xi / \omega_0$, and performing integrations over $y$ in Eq.(16) yields

$$F_z = \frac{\hbar\,\alpha(0)\omega_0}{4\pi^2 z_0^4}\int_0^{\infty} dx\, x^3 \frac{d^3 K_0(x)}{dx^3}\int_0^{\infty} dp \left[ \frac{\varepsilon(\mathrm{i}\, p\omega_0) - 1}{\varepsilon(\mathrm{i}\, p\omega_0) + 1} \right] \frac{(1 + p^2 - q^2 x^2)}{(1 + p^2 - q^2 x^2)^2 + 4p^2 q^2 x^2} + \\ + \frac{\hbar\,\alpha(0)\omega_0}{8\pi z_0^4}\int_{1/q}^{\infty} dx\, x^3 \frac{d^3 K_0(x)}{dx^3} \mathrm{Re}\left[ \frac{\varepsilon(\omega_0(qx-1)) - 1}{\varepsilon(\omega_0(qx-1)) + 1} \right], q = V/2\omega_0 z_0 \tag{17}$$

where $K_0(x)$ is a modified Bessel function. It is worthwhile noting the following useful relation

$$-d^3 K_0(x)/dx^3 = (1 + 2/x^2)K_1(x) + K_0(x)/x \tag{18}$$

A very compact formula stems from Eq.(17) if use is made of a nondissipative plasma model of a metallic half –space, $\varepsilon(\omega) = 1 - \omega_p^{\,2}/\omega^2$ ($\omega_p$ is the plasma frequency). Then the inner integral in the first term of Eq.(17) is calculated explicitly (see [8]) , and we get

$$F_z = \frac{\hbar\,\omega_s \alpha(0)}{8\pi z_0^4}\int_0^{\infty} x^3 \frac{d^3 K_0(x)}{dx^3}\left[ \frac{(1+\eta)\theta(1-qx)}{(1+\eta)^2 - q^2 x^2} + \frac{\theta(qx-1)}{\left[1 - (\eta + qx)^2\right]} \right] dx + \\ + \frac{\hbar\,\omega_0 \alpha(0)}{8\pi z_0^4}\eta^2 \int_{1/q}^{\infty} \frac{d^3 K_0(x)}{dx^3} \frac{x^3}{\eta^2 - (1-qx)^2} dx, \eta = \omega_s / \omega_0, \ \omega_s = \omega_p / \sqrt{2} \tag{19}$$

The condition of finite $\eta$ has been considered in [8]. For an ideal metal plate one should take the limit $\eta \to \infty$ . In this case Eq.(19) reduces to

$$F_z = -\frac{3\hbar\,\omega_0\alpha(0)}{8\,z_0^{\;4}} - \frac{\hbar\,\omega_0\alpha(0)}{8\,\pi\,z_0^{\;4}}\int\limits_{1/q}^{\infty} dx x^3\left(-\frac{d^3K_0(x)}{dx^3}\right) \tag{20}$$

At $q >> 1$, i.e. when $z_0 << V/2\omega_0$, that involves extremely small atom –surface separations since $V << c$, we get $3\pi$ for the integral, and the resulting $F_z$ is twice the first term in Eq.(20),

$$F_z \cong -\frac{3\hbar\omega_0\alpha(0)}{4z_0^{\;4}},\; z_0 << V/2\omega_0 \tag{21}$$

In the opposite case, at $q << 1$, on using

$$K_0(x) \cong \sqrt{\pi/2x}\,\exp(-x), \tag{22}$$

we get

$$F_z \cong -\frac{3\hbar\,\omega_0\alpha(0)}{8z_0^{\;4}} - \frac{3\hbar\,\omega_0\alpha(0)}{2\sqrt{\pi}\,z_0^{\;4}}\left(\frac{\omega_0 z_0}{V}\right)^{5/2}\exp(-2\omega_0 z_0/V),\; z_0 >> V/2\omega_0 \tag{23}$$

Specifically, the second term of Eq.(23) is responsible for the nonmonotonous force –velocity dependence.

### *2.2 An ideally conducting cavity wall*

In the case of an ideally conducting (reflecting) cavity wall, at $\varepsilon(\omega) \to \infty$, from Eqs. (6) it follows $\Delta_e(\omega) = 1,\; \Delta_m(\omega) = -1$, and Eqs.(14),(15) result in

$$F_z^{(0)} = -\frac{\hbar}{\pi^2\gamma}\int\limits_{-\infty}^{+\infty} dk_x \int\limits_{-\infty}^{+\infty} dk_y \int\limits_0^{\infty} d\xi\,(k^2 + \xi^2/c^2)\exp(-2\sqrt{k^2+\xi^2/c^2}\,z_0)\,\mathrm{Im}\left[\mathrm{i}\,\alpha(\gamma(\mathrm{i}\xi + k_x V))\right] \tag{24}$$

$$F_z^{(1)} = \frac{4\hbar}{\pi^2\gamma}\int\limits_0^{\infty} dk_x \int\limits_0^{\infty} dk_y \int\limits_0^{\infty} d\omega\,\alpha''(\gamma(\omega - k_x V))\,\mathrm{Re}\left[(k^2 - \omega^2/c^2)\exp(-2\sqrt{k^2-\omega^2/c^2}\,z_0)\right] \tag{25}$$

Substitutions into Eqs.(24), (25) via Eqs.(10) , (11) lead to ( $\lambda = 2\tilde{\omega}_0 z_0/c = 2\omega_0 z_0/\gamma c \equiv \lambda_0/\gamma$ )

$$F_z^{(0)} = \frac{4\hbar\alpha(0)\omega_0^5}{\pi^2 c^4\gamma^6}\int_0^\infty dx\int_0^\infty dy \frac{(1-\beta^2x^2+y^2)(x^2+y^2)^{3/2}}{\left[(1+\beta x)^2+y^2\right]\left[(1-\beta x)^2+y^2\right]}\left[\frac{d^3K_0(t)}{dt^3}\right]_{t=\lambda\sqrt{x^2+y^2}} \tag{26}$$

$$F_z^{(1)} = \frac{\hbar\omega_0\,\alpha(0)}{8\pi z_0^4\gamma}\int_{\lambda/\beta}^\infty \frac{dx x^4}{\sqrt{\lambda_0^2+x^2}}\frac{d^3K_0(x)}{dx^3} \tag{27}$$

Note that in deriving Eq.(27) no additional assumptions is used.

Eq.(26) can be simplified further in the limit cases $\beta\to 0$ and $\beta\to 1$ introducing the polar coordinates $x = r\cos\theta,\ y = r\sin\theta$ . Then, expanding the angular-dependent part of Eq. (26) with respect to $\beta$ and $(1-\beta)=1/2\gamma^2$ yields

$$\begin{aligned} A(x,y) = A(r,\theta) &\equiv \frac{1-\beta^2x^2+y^2}{\left[(1+\beta x)^2+y^2\right]\left[(1-\beta x)^2+y^2\right]} = \\ &= \frac{1-\beta^2r^2\cos^2\theta+r^2\sin^2\theta}{\left[(1+\beta r\cos\vartheta)^2+r^2\sin^2\theta\right]^2\left[(1-\beta r\cos\vartheta)^2+r^2\sin^2\theta\right]^2} = \\ &= \begin{cases} \dfrac{1}{1+\mathrm{r}^2\sin^2\theta} - \dfrac{\mathrm{r}^2\cos^2\theta(3r^2\sin^2\theta-1)}{(1+\mathrm{r}^2\sin^2\theta)^3}\beta^2, \beta<<1 \\ \dfrac{1-r^2\cos 2\theta}{(1+r^2)^2-4r^2\cos^2\theta} - \dfrac{1}{\gamma^2}\dfrac{r^2\cos^2\theta(1-2r^2-r^4+2r^4\cos 2\theta)}{((1+r^2)^2-4r^2\cos^2\theta)^2}, \gamma>>1 \end{cases} \end{aligned} \tag{28}$$

Inserting Eqs.(28) into Eq.(26) and using angular integrals ( $\theta(x)$ is the unit step-function)

$$\int_0^{\pi/2} A(r,\theta)d\theta = \begin{cases} \dfrac{\pi}{2\sqrt{1+r^2}} + \dfrac{\pi r^2}{4(1+r^2)^{3/2}}\beta^2, \beta<<1(\gamma\to 1) \\ \dfrac{\pi}{2}\theta(1-r)(1-r^2/2\gamma^2), \gamma>>1(\beta\to 1) \end{cases}, \tag{29}$$

yields

$$F_z^{(0)} \cong \begin{cases} \dfrac{\hbar\omega_0\alpha(0)}{8\pi z_0^4}\displaystyle\int_0^\infty dx \frac{x^4}{\sqrt{\lambda^2+x^2}}\frac{d^3K_0(x)}{dx^3}\left[1-\frac{(x^2+2\lambda^2)\beta^2}{2(\lambda^2+x^2)}\right], \beta<<1 \\ \dfrac{\hbar c\alpha(0)}{16\pi z_0^5\gamma}\displaystyle\int_0^\lambda dx x^4\frac{d^3K_0(x)}{dx^3}(1-x^2/2\lambda_0^2), \gamma>>1 \end{cases} \tag{30}$$

It is not difficult to get from Eqs. (27),(30) several simpler asymptotic formulas at different relations between parameters $\lambda_0$ , $\beta$ and $\gamma$ . These results are summarized in Table 1, being normalized by a factor $\hbar\omega_0\alpha(0)/z_0^4$ .

Table 1 Comparison of different force asymptotics

| No | Range | $F_z^{(0)}$ | No | Range | $F_z^{(1)}$ |
|---|---|---|---|---|---|
| 1 | $\lambda_0 << 1, \beta << 1$ | $-\frac{3}{8}\left(1-\frac{\beta^2}{2}\right)$ | 5 | $\lambda_0 << \beta << 1$ | $-\frac{3}{8}\left(1-\frac{\beta^2}{2}\right)$ |
| 2 | $\lambda_0 >> 1, \beta << 1$ | $-\frac{3}{\pi\lambda_0}\left(1-\frac{\beta^2}{2}\right)$ | 6 | $\beta << \lambda_0 << 1$ | $-\frac{1}{8\sqrt{2\pi}}\frac{\lambda_0^{5/2}}{\beta^{5/2}}\exp(-\lambda_0/\beta)$ |
| 3 | $\lambda_0 >> \gamma >> 1$ | $-\frac{3}{\pi}\frac{1}{\lambda_0\gamma}\left(1-\frac{10}{\lambda_0^2}\right)$ | 7 | $\lambda_0 >> 1 >> \beta$ | $-\frac{1}{8\sqrt{2\pi}}\frac{\lambda_0^{5/2}}{\beta^{7/2}}\exp(-\lambda_0/\beta)$ |
| 4 | $\lambda_0 << \gamma, \gamma >> 1$ | $-\frac{\lambda_0}{8\pi\gamma^3}$ | 8 | $\lambda_0 << 1 << \gamma$ | $-\frac{3}{8\gamma}$ |
| | | | 9 | $\lambda_0 >> \gamma >> 1$ | $-\frac{1}{8\sqrt{2\pi}}\frac{\lambda_0^{5/2}}{\gamma^{9/2}\beta^{7/2}}\exp(-\lambda_0/\gamma\beta)$ |
| | | | 10 | $1 << \lambda_0 << \gamma$ | $-\frac{3}{\pi\lambda_0\gamma}$ |

One can note complete agreement between Eq. (21) and Eqs. 1,5 of the table at $\beta << 1, \lambda_0 << 1$, with an obvious exception for relativistic corrections $\sim \beta^2$. In the case of large distances $z_0 >> V/2\omega_0$, which is equivalent to $\lambda_0 >> 1$, $\beta << 1$, the first term in Eq.(23) corresponding to $F_z^{(0)}$, differs from that in the table (Eq. 2), describing the Casimir –Polder force with first dynamical correction

$$F_z^{(0)} = -\frac{3}{2\pi}\frac{\hbar c\alpha(0)}{z_0^5}\left(1-\beta^2/2\right),\; z_0 >> V/2\omega_0 \tag{31}$$

This is not suprising because the abolimit involves a crossover with the atom –surface separation range where the retardation effects are of crucial importance.The dynamical correction in Eq. (31), as we see, has a character of repulsive force. This resembles the result of Barton [11] obtained for the image-charge force acting on a moving charged particle.

The nonrelativistic expression for $F_z^{(1)}$, corresponding to Eq. (23), is identical to Eq.6 in Table 1 (again with the exception for $\beta^2$ – correction, omitted in Eq. 6). Fig. 2(a,b) compares the normalized forces computed from Eqs. (27),(30) with their asymptotics at $\beta << 1$ and $\gamma >> 1$, as given in Table 1.

*2.3 Drude approximation for the plate material, general case*

The question of interest is how important is the influence of material parameters on the dynamical Casimir –Polder force. We now assume a half –space to be metallic, considering the dielectric permittivity within the Drude model approach with parameters of gold ($\omega_p = 1.37 \cdot 10^{16}\ rad/s$, $\tau = 1.89 \cdot 10^{-14}\ s$)

$$\varepsilon(\omega) = 1 - \frac{\omega_p{}^2}{\omega(\omega + \mathrm{i}/\tau)} \tag{32}$$

Eq.(14) is explicitly reduced to a double integral using the polar coordinates $(k,\theta)$ instead $(k_x, k_y)$ and performing integration over $\theta$. This is done with allowance for the identities ($a \equiv kV\cos\theta$)

$$\frac{\tilde{\omega}_0{}^2 + \xi^2 - a^2}{\left[(\tilde{\omega}_0 + a)^2 + \xi^2\right]\left[(\tilde{\omega}_0 - a)^2 + \xi^2\right]} \equiv \frac{1}{4\tilde{\omega}_0}\left[\frac{1}{\tilde{\omega}_0 + \mathrm{i}\xi - a} + \frac{1}{\tilde{\omega}_0 - \mathrm{i}\xi - a} + \frac{1}{\tilde{\omega}_0 + \mathrm{i}\xi + a} + \frac{1}{\tilde{\omega}_0 - \mathrm{i}\xi + a}\right] \tag{33}$$

$$\frac{1}{\left[(\tilde{\omega}_0 + a)^2 + \xi^2\right]\left[(\tilde{\omega}_0 - a)^2 + \xi^2\right]} \equiv -\frac{\mathrm{i}}{8\xi\tilde{\omega}_0}\left[\frac{1}{(\tilde{\omega}_0 - \mathrm{i}\xi)}\left[\frac{1}{\tilde{\omega}_0 - \mathrm{i}\xi + a} + \frac{1}{\tilde{\omega}_0 - \mathrm{i}\xi - a}\right] - \frac{1}{(\tilde{\omega}_0 + \mathrm{i}\xi)}\left[\frac{1}{\tilde{\omega}_0 + \mathrm{i}\xi + a} + \frac{1}{\tilde{\omega}_0 + \mathrm{i}\xi - a}\right]\right] \tag{34}$$

Bearing in mind Eqs.(33), (34) and making use transformations of the integrand variables $x = \omega/\tilde{\omega}_0$, $k^2c^2/\omega^2 = u^2 + 1$, Eq.(14) is rationalized in the simple form

$$F_z^{(0)} = -\frac{\hbar\omega_0\,\alpha(0)}{64\pi^2 z_0{}^4}\frac{1}{\lambda}\int_1^\infty \frac{du}{u^4}\int_0^\infty dt\, t^4 \phi(t/\lambda u, u, \beta)\exp(-t) \tag{35}$$

$$\phi(x,u,\beta) = \sum_{n=1}^{3} A_n(x,u,\beta) Q_n(x,u,\beta) \tag{36}$$

$$A_1(x,u,\beta) = \Delta_e(u,x)\left[(2-\beta^2)(u^2-1) + (1-2\beta^2)\right] + \Delta_m(u,x)\left[\beta^2(u^2-1) - 1\right] \tag{37}$$

$$A_2(x,u,\beta) = (u^2+1)\left[\Delta_e(u,x) + \Delta_m(u,x)\right] \tag{38}$$

$$A_3(x,u,\beta) = 2\,\mathrm{i}\,\beta^2 x(u^2-1)\left[\Delta_e(u,x) + \Delta_m(u,x)\right] \tag{39}$$

$$\begin{aligned} Q_1(x,u,\beta) = {} & f_0(1+\mathrm{i}\,x, \beta x\sqrt{u^2-1}) + f_0(1-\mathrm{i}\,x, \beta x\sqrt{u^2-1}) + f_0(1+\mathrm{i}\,x, -\beta x\sqrt{u^2-1}) + \\ & + f_0(1-\mathrm{i}\,x, -\beta x\sqrt{u^2-1}) \end{aligned} \tag{40}$$

$$\begin{aligned} Q_2(x,u,\beta) = {} & f_2(1+\mathrm{i}\,x, \beta x\sqrt{u^2-1}) + f_2(1-\mathrm{i}\,x, \beta x\sqrt{u^2-1}) + f_2(1+\mathrm{i}\,x, -\beta x\sqrt{u^2-1}) + \\ & + f_2(1-\mathrm{i}\,x, -\beta x\sqrt{u^2-1}) \end{aligned} \tag{41}$$

$$\begin{aligned} Q_3(x,u,\beta) = {} & \frac{1}{(1-\mathrm{i}\,x)}\Big[f_0(1-\mathrm{i}\,x, \beta x\sqrt{u^2-1}) - f_2(1-\mathrm{i}\,x, \beta x\sqrt{u^2-1}) + \\ & + f_0(1-\mathrm{i}\,x, -\beta x\sqrt{u^2-1}) - f_2(1-\mathrm{i}\,x, -\beta x\sqrt{u^2-1})\Big] - \frac{1}{(1+\mathrm{i}\,x)}\Big[f_0(1+\mathrm{i}\,x, \beta x\sqrt{u^2-1}) - \\ & - f_2(1+\mathrm{i}\,x, \beta x\sqrt{u^2-1}) + f_0(1+\mathrm{i}\,x, -\beta x\sqrt{u^2-1}) - f_2(1+\mathrm{i}\,x, -\beta x\sqrt{u^2-1})\Big] \end{aligned} \tag{42}$$

where $\lambda = 2\tilde{\omega}_0 z_0 / c = 2\omega_0 z_0 / \gamma c = \lambda_0 / \gamma$ ,and the auxiliary functions $f_{0,2}(x, y)$ are given by

$$f_0(x,y) = \int_0^{\pi} \frac{dz}{x + y\cos z} = \frac{\pi}{\sqrt{x^2 - y^2}}\,\chi\Big[(x-y)\Big(\mathrm{Re}\sqrt{x^2-y^2} - \mathrm{i}\cdot\mathrm{Im}\sqrt{x^2-y^2}\Big)\Big] \tag{43}$$

$$f_2(x,y) = \int_0^{\pi} \frac{\sin^2 z\, dz}{x + y\cos z} = \frac{\pi\, x}{y^2}\left(1 - \sqrt{1 - y^2/x^2}\right) \tag{44}$$

$$\chi(x) = \begin{cases} 1, \mathrm{Re}(x) > 0 \\ 1, \mathrm{Re}(x) = 0, \mathrm{Im}(x) > 0 \\ -1, \mathrm{Re}(x) < 0 \\ -1, \mathrm{Re}(x) = 0, \mathrm{Im}(x) < 0 \end{cases} \tag{45}$$

In Eqs.(30)-(32), according to the used notations, the reflection coefficients $\Delta_{e,m}(\omega)$ (see Eqs.(4)) take the form

$$\Delta_e(u,x) = \frac{u\varepsilon(\mathrm{i}x) - \sqrt{u^2 + \varepsilon(\mathrm{i}x) - 1}}{u\varepsilon(\mathrm{i}x) + \sqrt{u^2 + \varepsilon(\mathrm{i}x) - 1}}\ ,\ \Delta_m(u,x) = \frac{u - \sqrt{u^2 + \varepsilon(\mathrm{i}x) - 1}}{u + \sqrt{u^2 + \varepsilon(\mathrm{i}x) - 1}} \tag{46}$$

One should stress that, despite its complex form, the integrand function in Eq.(35) proves to be purely real.

Eq.(13), in its turn, is rationalized if use is made of the dimensionless variables $x = 2z_0 k_x$ , $y = 2z_0 k_y$ :

$$F_z^{(1)} = -\frac{\hbar\omega_0\,\alpha(0)}{16\pi\, z_0^{\;4}} \int\limits_{\lambda/\beta}^{\infty} dx \int\limits_{0}^{\infty} dy\, \mathrm{Re}\left[\exp\left(-\sqrt{x^2 + y^2 - (\lambda - \beta x)^2}\right) B(x, y, \beta, \lambda)\right] \tag{47}$$

$$B(x, y, \beta, \lambda) = \left[2\left(x^2 + y^2 - \beta^2 x^2\right)\Delta_e(x, y, \beta, \lambda) + 2\beta^2 y^2 \Delta_m(x, y, \beta, \lambda)\right]\left(1 - \frac{(\lambda - \beta x)^2}{x^2 + y^2}\right) + \\ + \lambda^2\left[\Delta_e(x, y, \beta, \lambda) + \Delta_m(x, y, \beta, \lambda)\right] \tag{48}$$

$$\Delta_e(x, y, \beta, \lambda) = \frac{\sqrt{x^2 + y^2 - (\lambda - \beta x)^2}\,\varepsilon(-\tilde{\omega}_0^{\;-}) - \sqrt{x^2 + y^2 - (\lambda - \beta x)^2\,\varepsilon(-\tilde{\omega}_0^{\;-})}}{\sqrt{x^2 + y^2 - (\lambda - \beta x)^2}\,\varepsilon(-\tilde{\omega}_0^{\;-}) + \sqrt{x^2 + y^2 - (\lambda - \beta x)^2\,\varepsilon(-\tilde{\omega}_0^{\;-})}} \tag{49}$$

$$\Delta_m(x, y, \beta, \lambda) = \frac{\sqrt{x^2 + y^2 - (\lambda - \beta x)^2} - \sqrt{x^2 + y^2 - (\lambda - \beta x)^2\,\varepsilon(-\tilde{\omega}_0^{\;-})}}{\sqrt{x^2 + y^2 - (\lambda - \beta x)^2} + \sqrt{x^2 + y^2 - (\lambda - \beta x)^2\,\varepsilon(-\tilde{\omega}_0^{\;-})}} \tag{50}$$

$$\tilde{\omega}_0^{\;-} = \tilde{\omega}_0(1 - \beta\, x/\lambda) = \frac{c}{2z_0}(\lambda - \beta x), \;\; \beta = V/c \tag{51}$$

It is not difficult to see that for an ideal metal plate, $\Delta_e = 1, \Delta_m = -1$ Eqs.(35),(47) exactly reproduce Eqs.(14),(15). At this point, we want to note that the parametrization used is not a unique, but it seems to be an optimal, since it allows to get rapid convergence of the involved double integrals when making numerical calculations.

## 3. Numerical results

To clearly demonstrate results of Section 2.3, for a numerical example we have chosen parameters of ground state $Cs$ atoms ($\alpha(0) = 57 \cdot 10^{-30}\, m^3, \omega_0 = 1.44\, eV$ [14]), and of gold for the surface (Eq. (32)).

Fig. 3 and Fig. 4 compare the force components $F_z^{(0)}$, $F_z^{(1)}$ and their sum, computed in Drude approximation (a), and in case of an ideally conducting metal plate (b). All forces are normalized by a factor $\hbar\omega_0\alpha(0)/z_0^{\ 4}$ and are plotted with minus sign. The dashed, dotted and solid lines correspond to $F_z^{(1)}, F_z^{(0)}$ , and total force $F_z^{(0)} + F_z^{(1)}$ , all being computed at $\lambda_0 = 7.5$ as functions of $\beta$ (Fig.3) and $\gamma$ (Fig. 4). One can note several interesting features in these figures. i) Appearance of local force minimum and maximum near $\gamma = 5$ and $\gamma = 10$ (in Drude case). ii) The component $F_z^{(1)}$ proves to be a dominating contribution at $\gamma > 5$, irrespectively of the involved approximation for $\varepsilon(\omega)$. iii) The forces calculated with allowance for material parameters in Drude case are larger by an order of magnitude than in case of an ideal plate material (at $\gamma > 10$). In addition, in the former case the Casimir –Polder force has a much longer tail. Contrary to that, the components $F_z^{(0)}$ calculated in both approximations for the plate material, are very similar (cf. the dotted lines in Fig.3(a,b), and dashed-dotted lines in Fig. 4(a,b)). With decreasing $\lambda_0$ (i.e. with decreasing separation distance), the extrema points are shifted to smaller $\beta$ values (see Fig. 5(a), $\lambda_0 = 0.75$), and further on, a lengthy plateau is formed in the velocity range $0.2 < \beta < 0.8$ ($\lambda_0 = 0.075$, Fig. 5(b)), where the force proves to be nearly $\beta -$independent. At $\beta < 0.2$ (Fig.5(b)), the force very rapidly increases and tends to the static values. Approximately the same features have been observed in calculations of the nonretarded dynamical van –der –Waals potential [8].

Interestingly, a qualitatively similar nonmonotonous velocity dependence of dynamical image-charge force acting on a relativistic charged particle moving near a dielectric surface was found in [15], where the authors related the force maximum with a manifestation of the Cherenkov effect at $\beta = 1/\sqrt{\varepsilon}$ . For metals, obviously, this condition is out of sense. Contrary to this, in our case, the nonmonotonous velocity dependence of the Casimir –Polder force is entirely caused by a nonlinear character of the component $F_z^{(1)}$, and an appreciable influence of material properties (of metal substrate), shifting the peak of the $F_z^{(1)}$ to larger $\beta$. For an ideally reflecting plate the second factor is absent, but the tendency to form the force maximum is clearly seen (see Figs. 3(b) and 4(b)). Thus, qualitatively, such a behavior of $F_z^{(1)}$ is similar to that governed by Eqs.6,7,9 of Table 1.

Finally, to see the difference between the Drude case and an ideally reflecting plate, we have plotted in Fig. 6 the dependences $F_z(z_0)$ calculated at different $\gamma$. The dashed lines correspond to $\gamma = 1$ (static Casimir –Polder force), the upper and lower solid lines –to $\gamma = 5,\ 50$ , and dotted

lines –to $\gamma = 500$ . Comparing Figs. 6(a),(b), we see that material parameters influence in such a manner, that the difference between the static forces and dynamical ones (irrespectively of $\gamma$) becomes much less. As a result, the dynamical forces turn out to be by 1 to 3 orders of magnitude larger than for an ideally reflecting plate (at $20 < z_0 < 500\,nm$ , in our case).

## 3. Conclusions

We have got closed integral expressions for the dynamical Casimir –Polder force applied to the ground state neutral atom moving parallel to a flat boundary of cavity wall. The cases of ideal and Drude-like metallic walls have been studied in detail. The relativistic formulas coincide with those obtained in the nonrelativistic and nonretarded approximation $\beta \to 0$ .

A noticeable point of the present analysis appears to be a complex dependence of the dynamical Casimir –Polder force on the velocity (energy), distance and material properties. In particular, the dependence on $\beta, \gamma$ proves to be nonmonotonous, showing an appearance of local maximum or a wide plateau when $\beta$ increases. At separation distances ranging from several tens to several hundreds *nm*, the dynamical forces for a metallic wall described by the Drude dielectric function turn out to be by 1 to 3 order of magnitude larger than in the case of an ideal metallic wall.

The obtained results can be useful for theoretical interpretation of the experiments on passage through cavities and surface reflection of subrelativistic neutral atomic beams. Experimentally one might consider passing of ions with small ionization degree through gas targets, electrostatically deflecting away from the beam those ions which do not capture an electron, and subsequently scattering the remaining neutral atoms near a surface.

# FIGURE CAPTIONS

Fig.1. Schematic of the atom –surface interaction.

Fig.2(a,b) Comparison of Eqs. (27), (30) with the asymptotic relations given in Table 1. The normalization of forces is with respect to $\hbar\omega_0\alpha(0)/z_0{}^4$. Case (a): upper solid line –Eq.(30), lower solid line –Eq.(27), dotted line –Eq.1 from Table 1, dashed line –Eq.2, dashed-dotted line –Eq.6. Case (b): upper solid line –Eq.(27), lower solid line –Eq.(30), dashed line –Eq.3 from Table 1, dotted line –Eq.10, dashed –dotted line –Eq.4.

Fig.3(a,b) Dependence of dynamical Casimir –Polder force on velocity for $Cs$ atom interacting with $Au$ plate (a), and with perfectly reflecting plate (b). All forces are given in normalized form similar to Fig.2; $\lambda_0 = 7.5$ ($z_0 = 500\,nm$). The dotted, dashed and solid lines correspond to Eqs.(35), (47) and their sum in case (a); in case (b) the dashed-dotted, dashed and solid lines correspond to Eqs.(27), (30) and their sum, the dotted line –to Eq.10 from Table 1.

Fig.4(a,b) The same as on Fig.3(a) with respect to energy: (a) $Au$ plate, (b) perfect plate.

Fig.5(a,b) The same as on Fig. 3(a) at $\lambda_0 = 0.75$(a) and $\lambda_0 = 0.075$(b).

Fig.6(a,b) Dependence of dynamical Casimir –Polder force on distance for $Cs$ atom interacting with $Au$ plate (a), and with perfectly reflecting plate (b). The normalization of forces is with respect to $\hbar\omega_0{}^5\alpha(0)/c^4$. The dashed lines –static case, $\gamma = 1$, upper solid lines $-\gamma = 5$, lower solid lines $-\gamma = 50$, dotted lines $-\gamma = 500$.

FIGURE 1

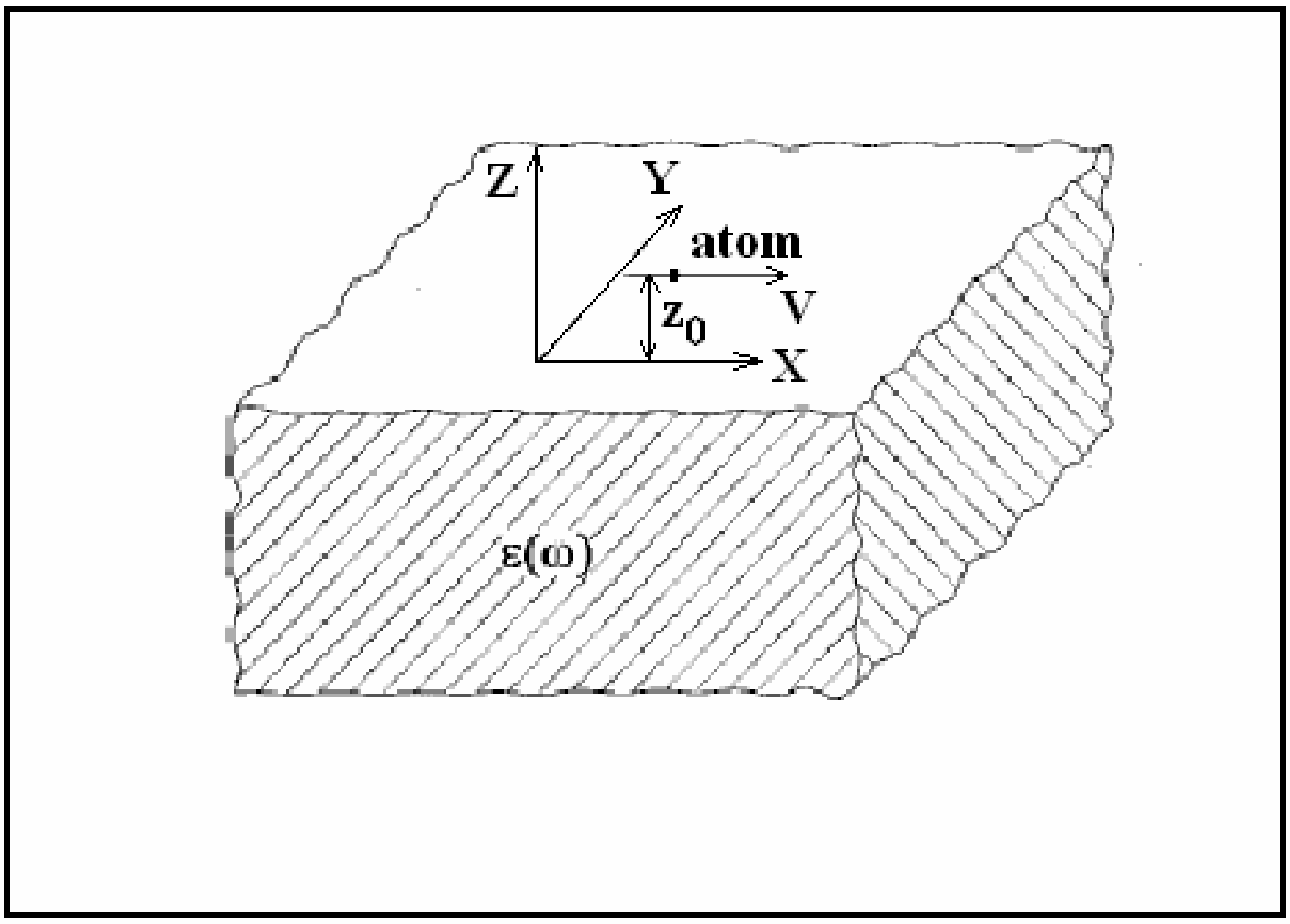

FIGURE 2(a)

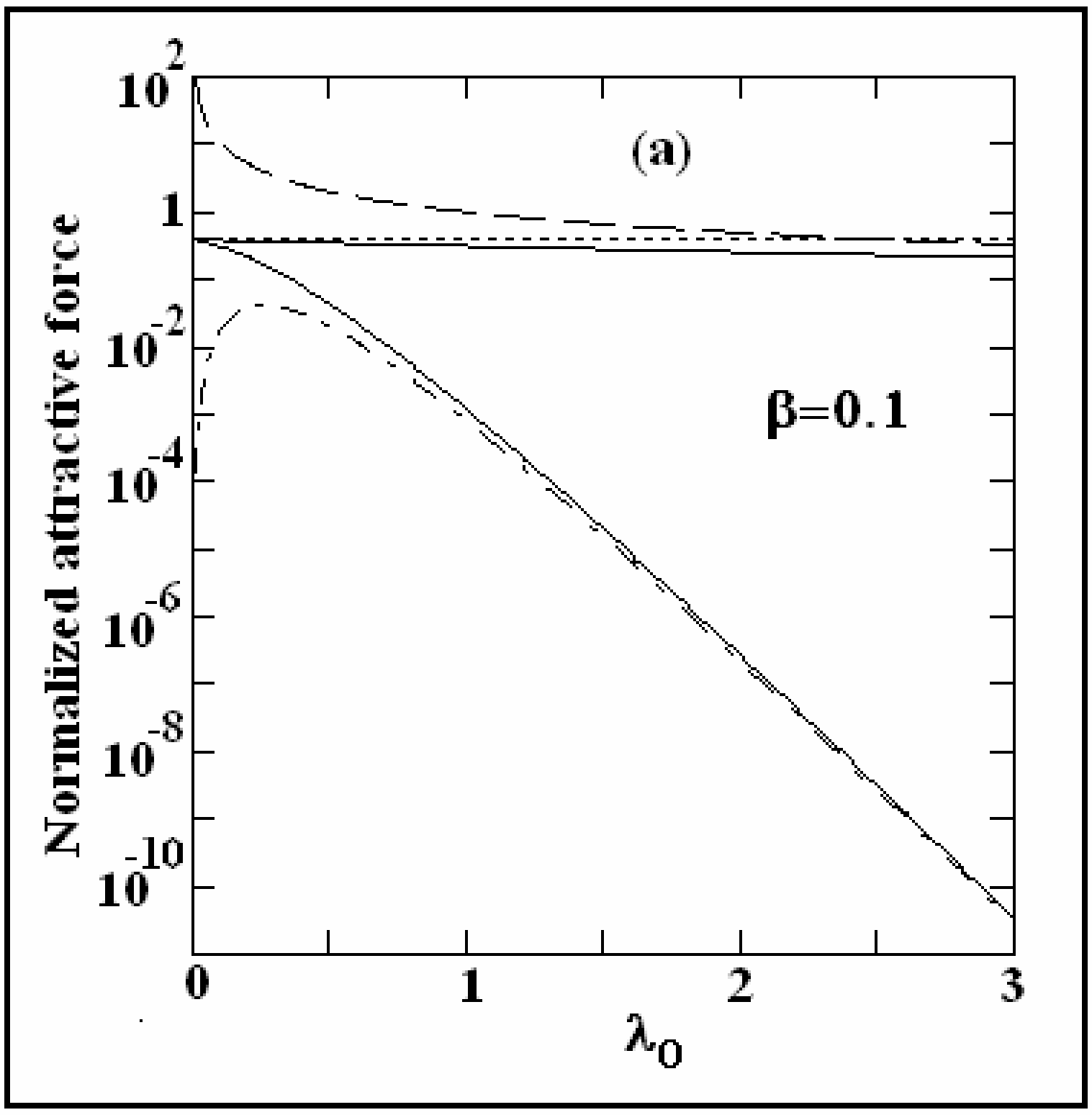

FIGURE 2(b)

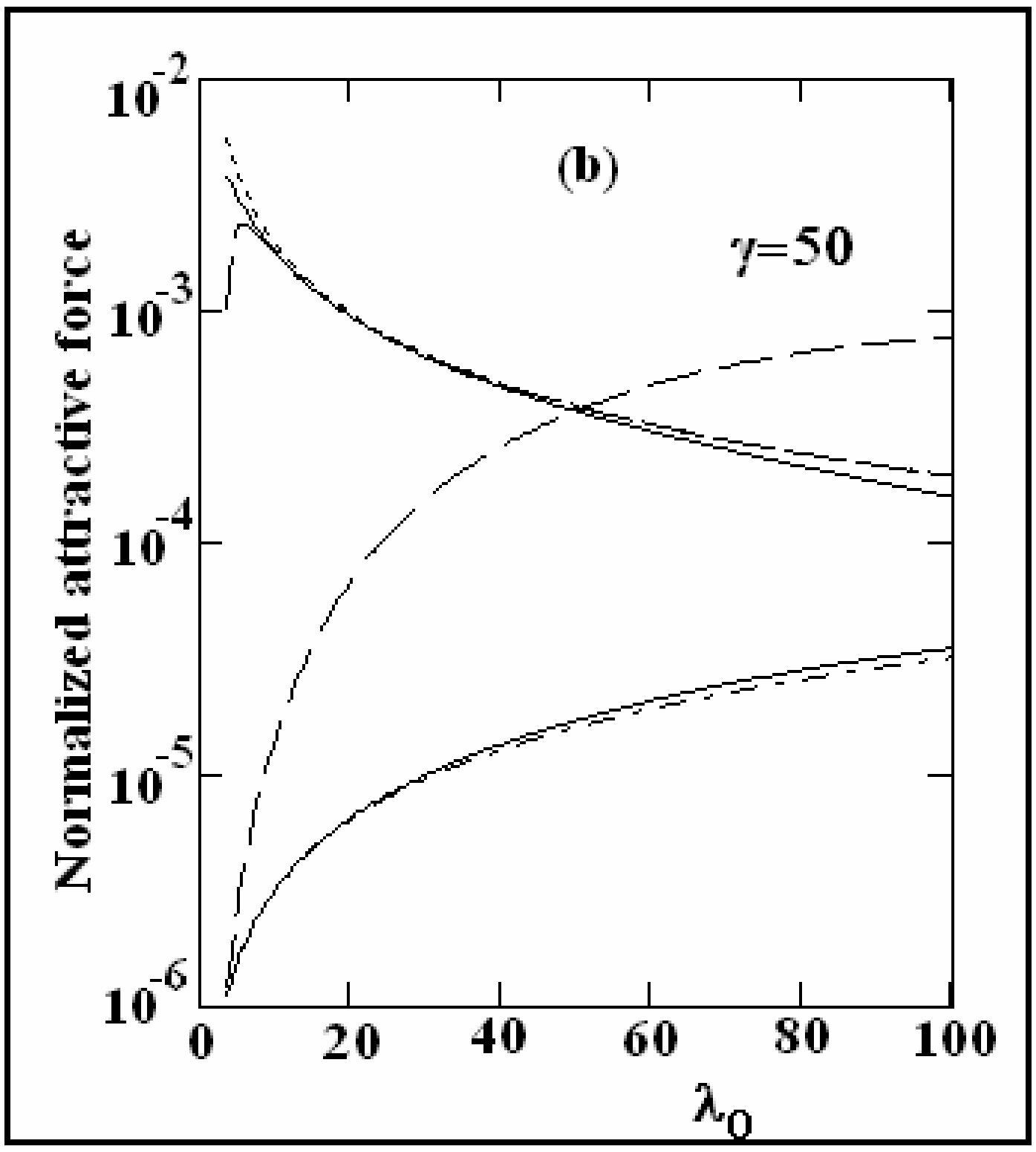

FIGURE 3(a)

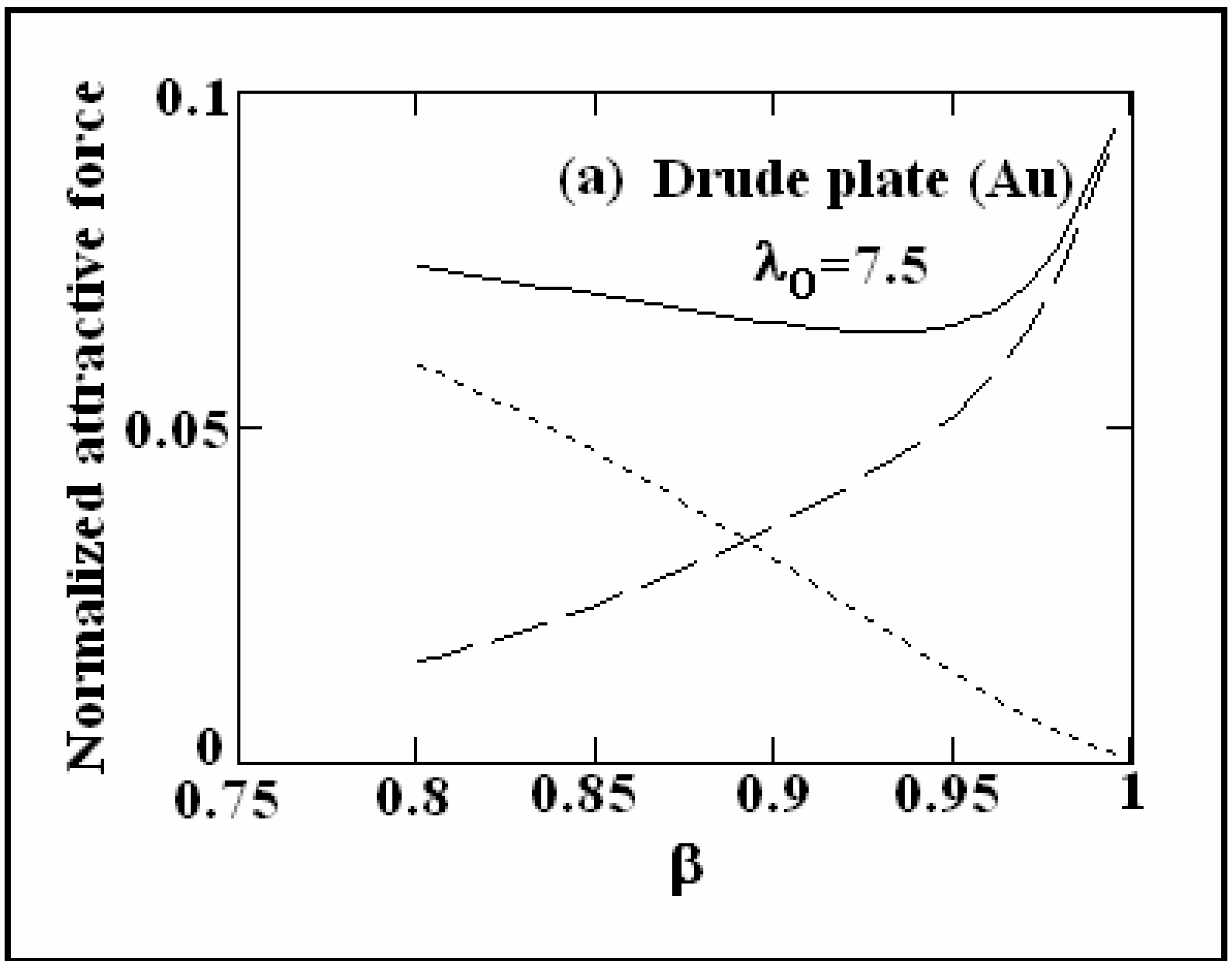

FIGURE 3(b)

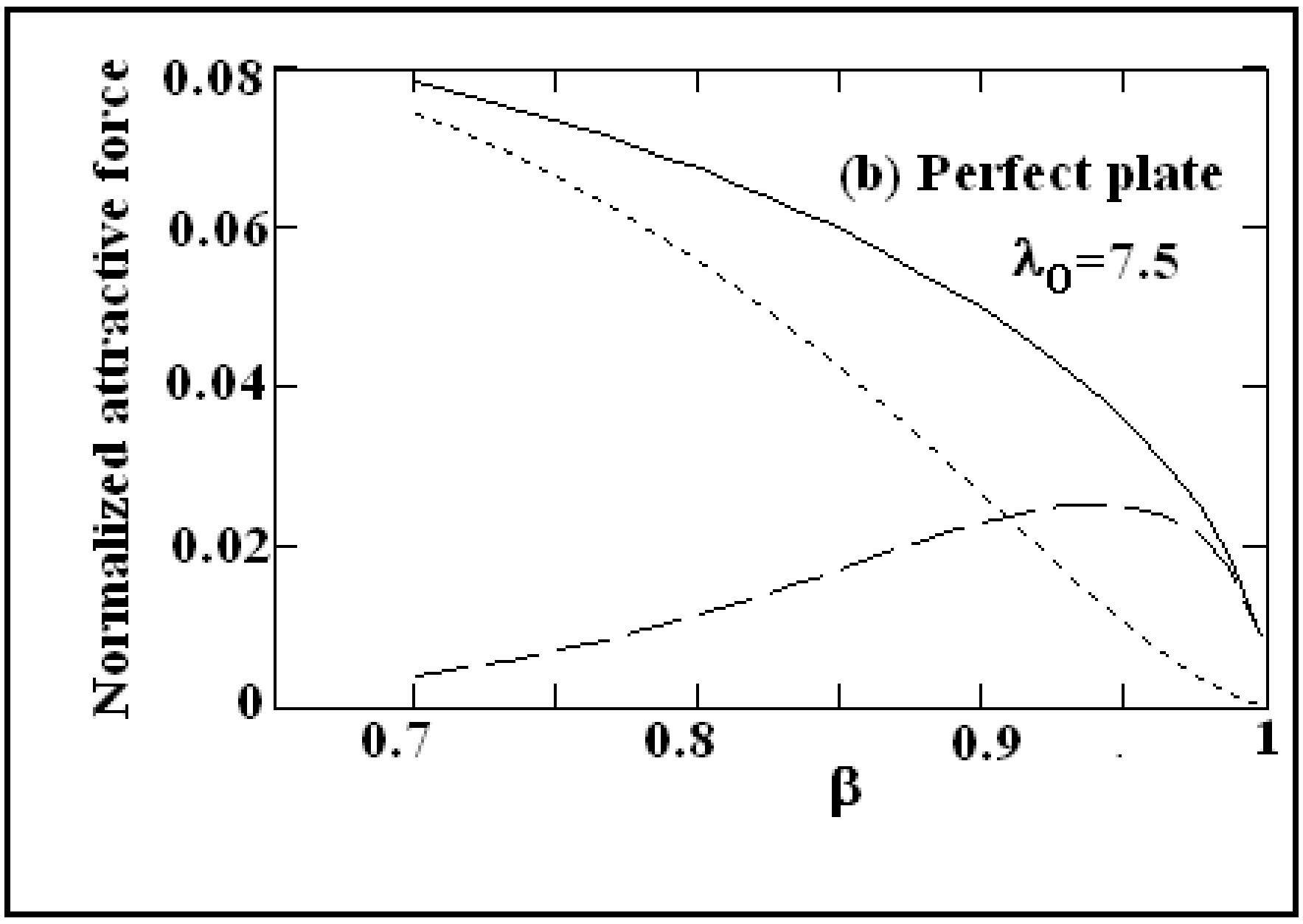

FIGURE 4(a)

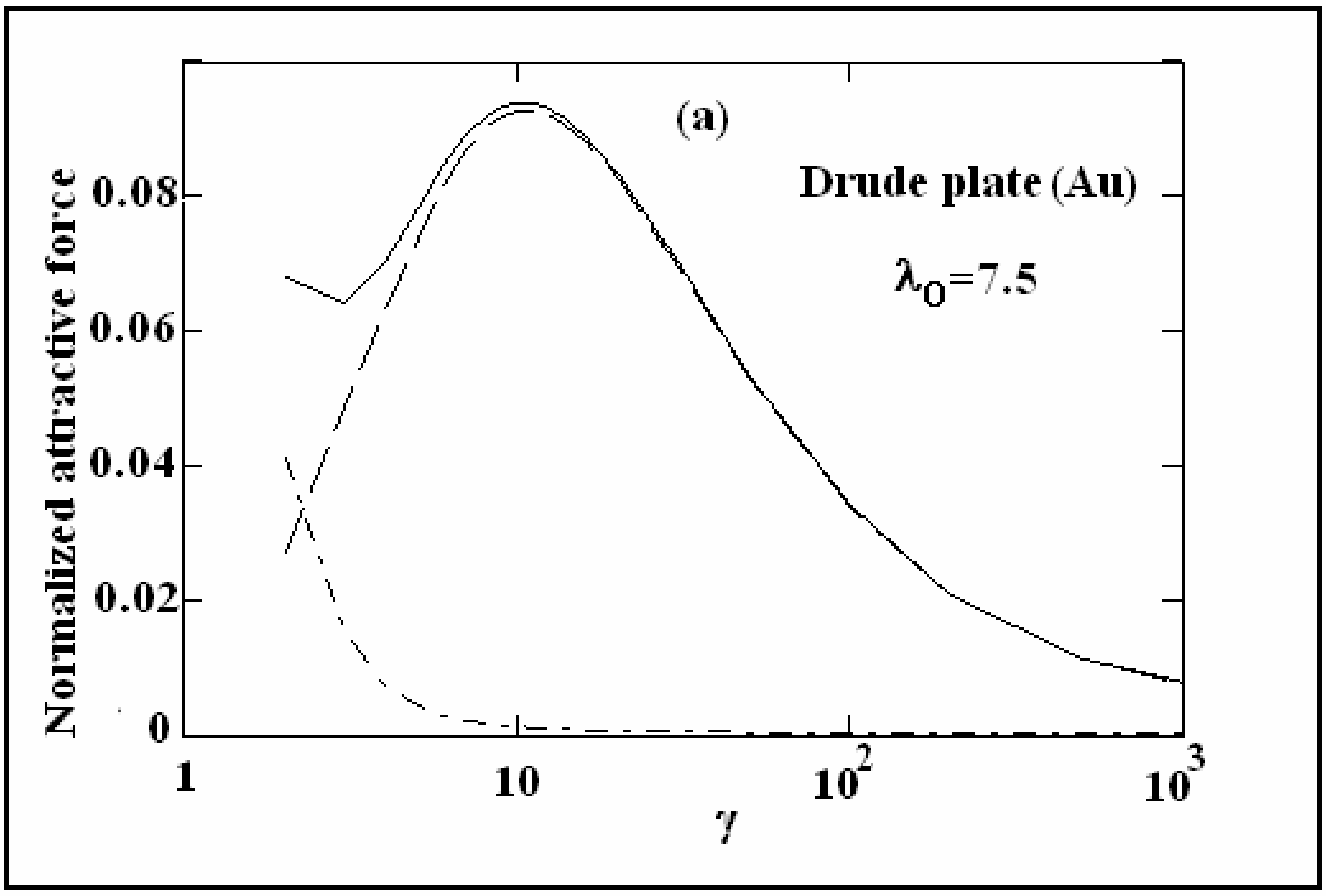

FIGURE 4(b)

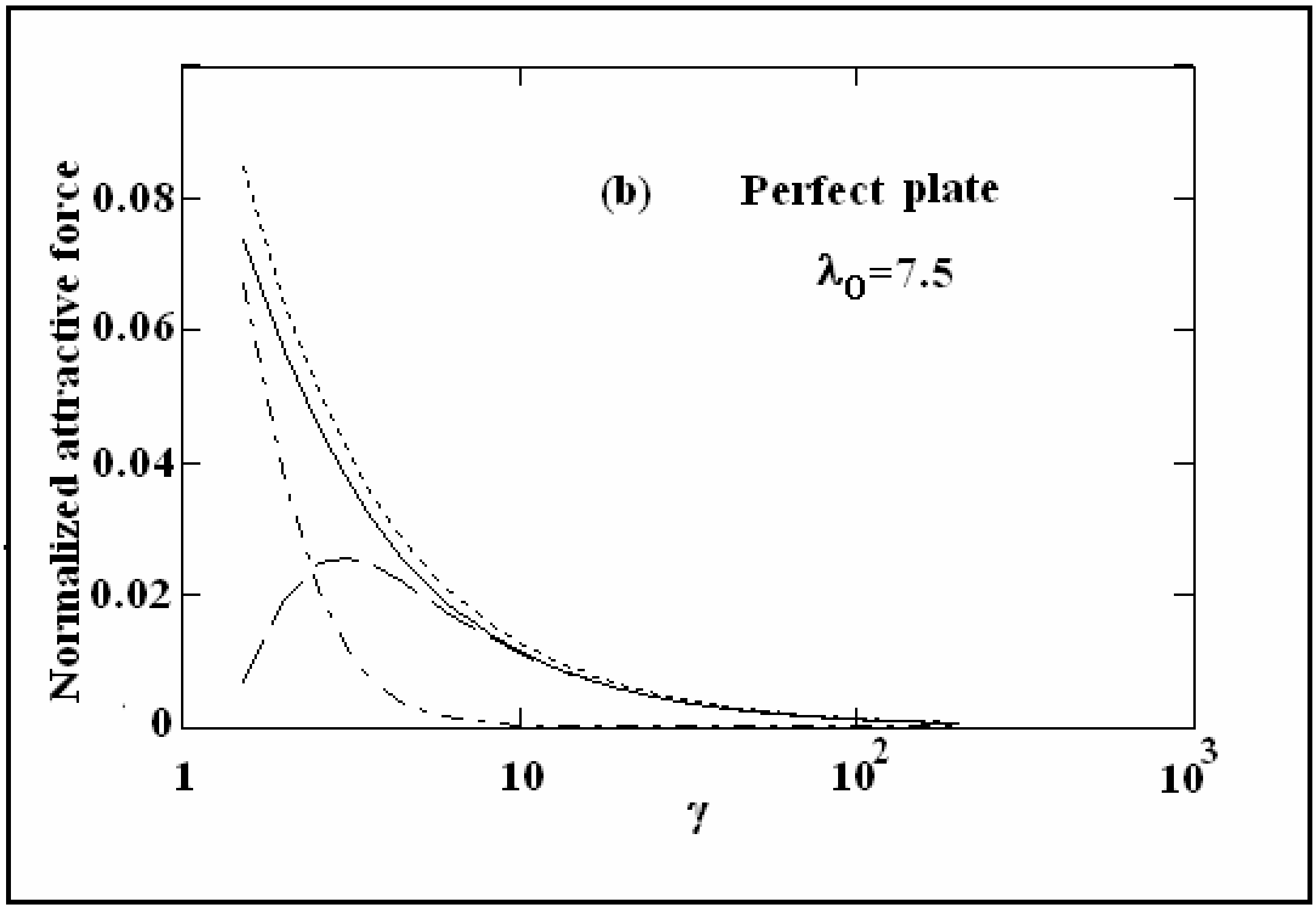

FIGURE 5(a)

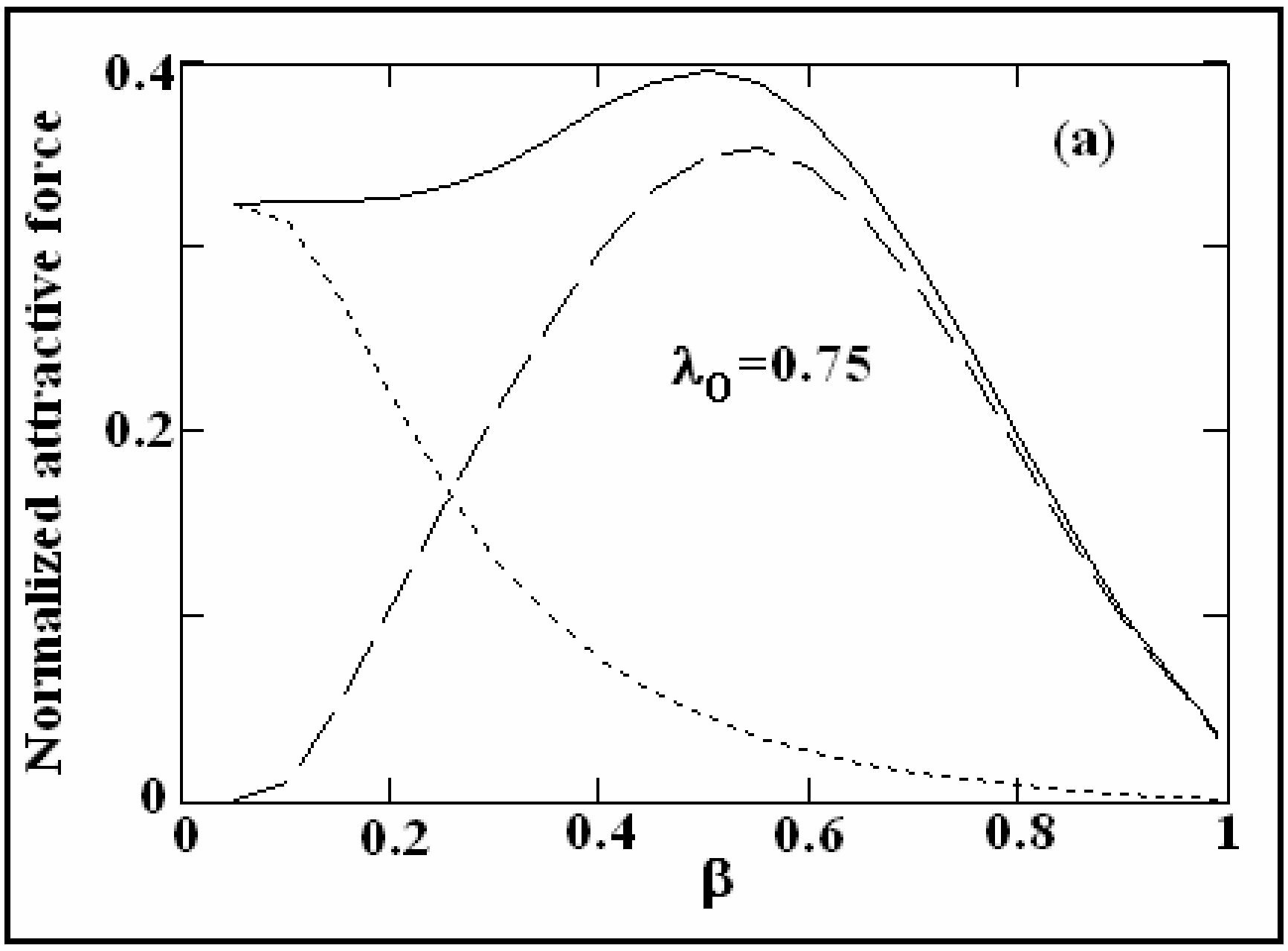

FIGURE 5(b)

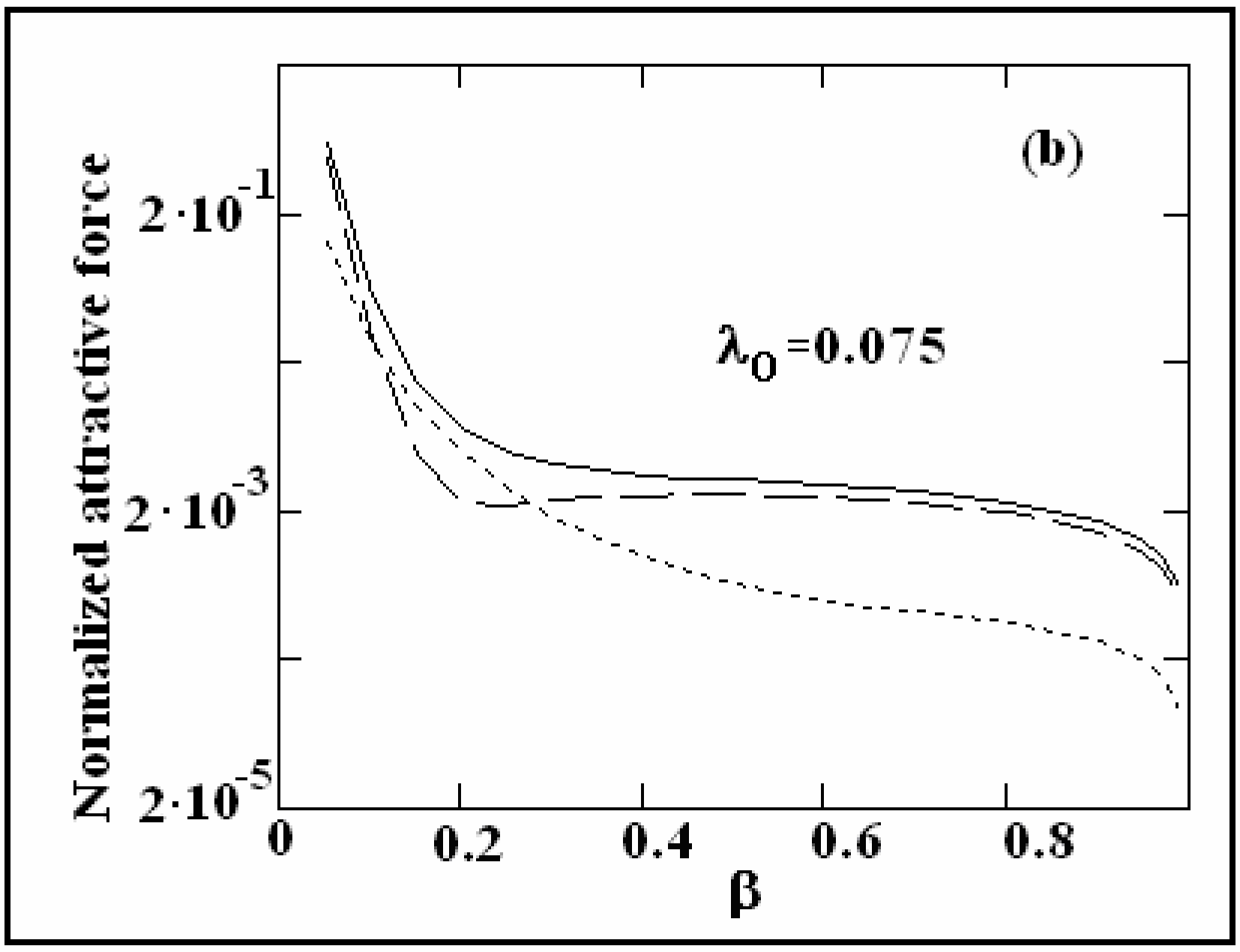

FIGURE 6(a)

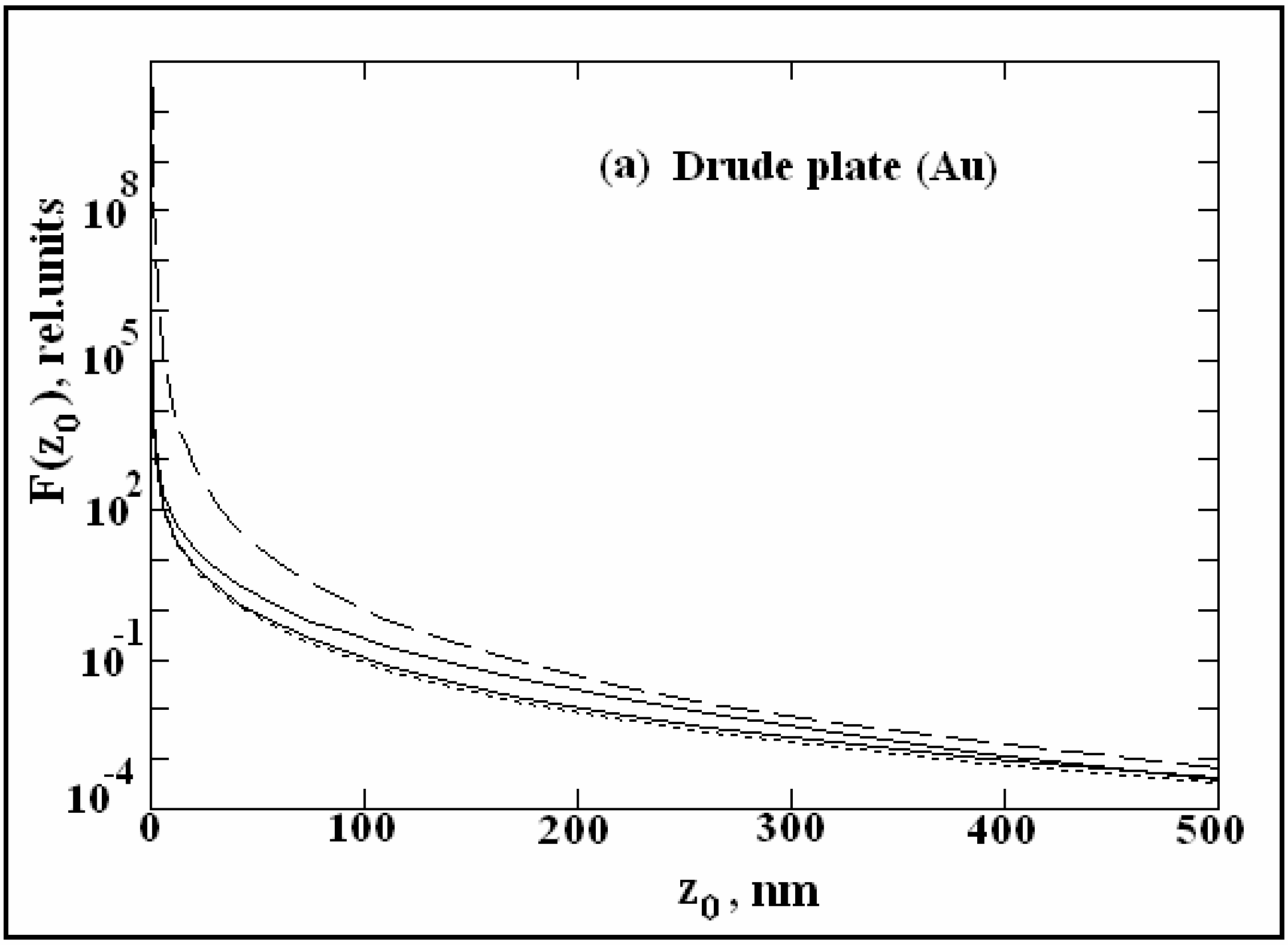

FIGURE 6(b)

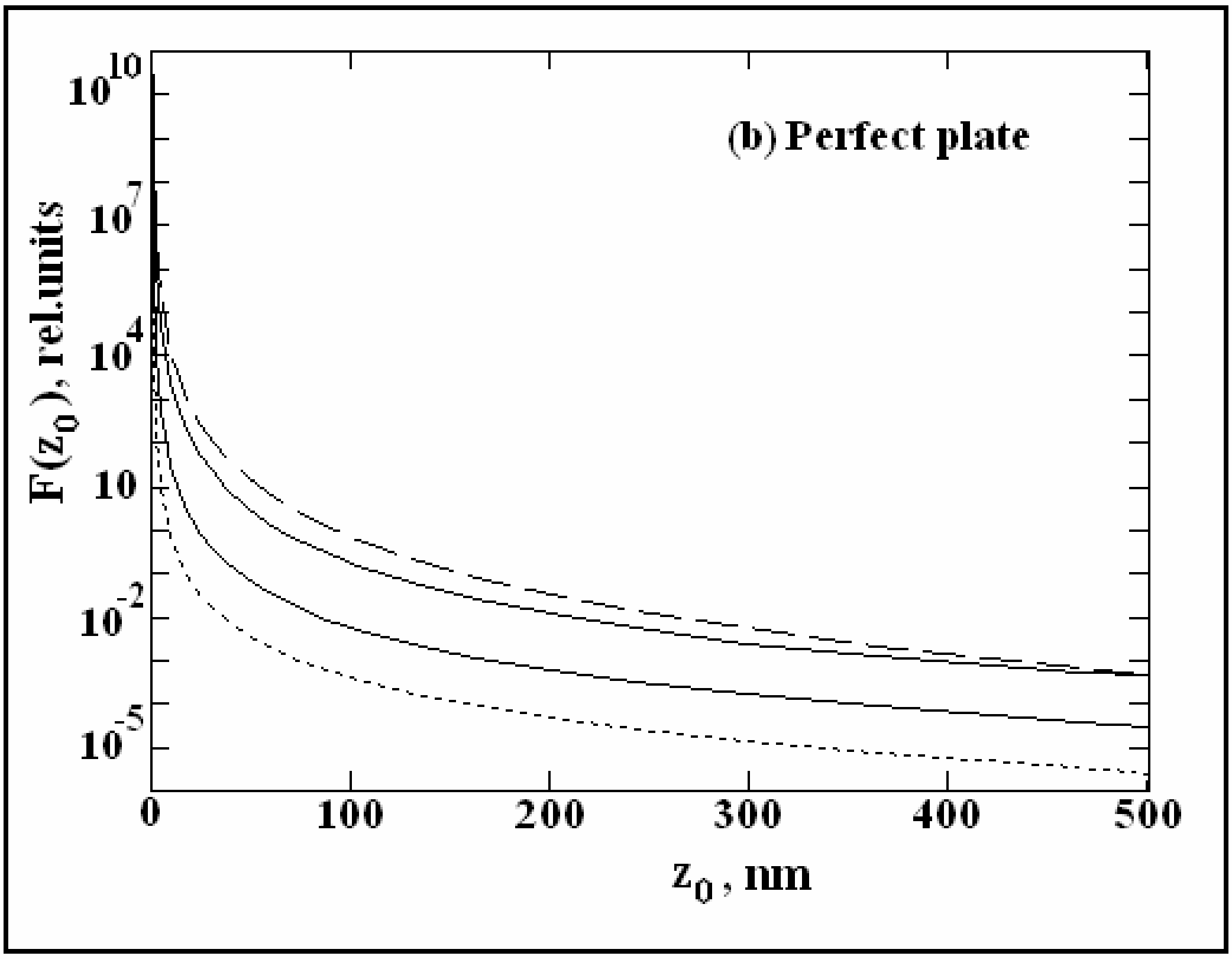